\documentclass[prl,preprint,superscriptaddress,longbibliography]{revtex4-1}

\usepackage[english]{babel}
\usepackage{lipsum}  

\usepackage{physics}
\usepackage{amsmath}
\usepackage{graphicx}
\usepackage{multirow}
\usepackage{longtable,threeparttablex}
\graphicspath{{figs/}}
\usepackage{hyperref}
\usepackage{footmisc}
\usepackage{bigints}
\hypersetup{
 bookmarks=true,		
 unicode=false,			
 pdftoolbar=true,		
 pdffitwindow=false,		
 pdfstartview={FitH},		
 pdfcreator={pdflatex},		
 pdfproducer={vim},		
 pdfkeywords={quantum chemistry} {chemical bond} {electron-deficit bonds} {quantum correlations} {entanglement} {multiorbital correlations} {diborane} {C-Be-C complex}, 
 pdfnewwindow=true,		
 colorlinks=true,		
 linktoc=page,			
 linkcolor=blue,		
 citecolor=blue,		
 filecolor=blue,		
 urlcolor=blue			
}
\usepackage{multirow}

\begin{document}

\title{Accurate Hyperfine Tensors for Solid State Quantum Applications: Case of the NV Center in Diamond}

\date{\today}
\author{Istv\'{a}n Tak\'{a}cs}
\affiliation{Department of Physics of Complex Systems, E\"otv\"os Lor\'and University, Egyetem t\'er 1-3, H-1053 Budapest, Hungary}
\affiliation{MTA–ELTE Lend\"{u}let "Momentum" NewQubit Research Group, P\'azm\'any P\'eter, S\'et\'any 1/A, 1117 Budapest, Hungary}

\author{Viktor Iv\'{a}dy}
\email{ivady.viktor@ttk.elte.hu}
\affiliation{Department of Physics of Complex Systems, E\"otv\"os Lor\'and University, Egyetem t\'er 1-3, H-1053 Budapest, Hungary}
\affiliation{MTA–ELTE Lend\"{u}let "Momentum" NewQubit Research Group, P\'azm\'any P\'eter, S\'et\'any 1/A, 1117 Budapest, Hungary}
\affiliation{Department of Physics, Chemistry and Biology, Link\"oping University, SE-581 83 Link\"oping, Sweden}

\begin{abstract}
The decoherence of point defect qubits is often governed by the electron spin-nuclear spin hyperfine interaction that can be parameterized by using ab inito calculations in principle. So far most of the theoretical works have focused on the hyperfine interaction of the closest nuclear spins, while the accuracy of the predictions for distinct nuclear spins is barely discussed. Here we demonstrate for the case of the NV center in diamond that the absolute relative error of the computed hyperfine parameters can exceed 100\% using an industry standards first-principles code. To overcome this issue, we implement an alternative method and report on significantly improved hyperfine values with $\mathcal{O}$(1\%) relative mean error at all distances. The provided accurate hyperfine data for the NV center enables high-precision simulation of NV quantum nodes for quantum information processing and positioning of nuclear spins by comparing experimental and theoretical hyperfine data. 
\end{abstract}

\maketitle

\section*{Introduction}

Point defects have been widely used to control the optical and electronic properties of semiconductors.  Recently, the magnetic properties of these materials have also been tailored by paramagnetic defects giving rise to various microscopic and mesoscopic magnetic phenomena. At low defect counteractions, controllable few-spin systems can be realized that has led to the development of point defect quantum bits\cite{jelezko_spectroscopy_2001,weber_quantum_2010} (qubits) and quantum nodes\cite{wehner_quantum_2018,bradley_ten-qubit_2019,abobeih_fault-tolerant_2022}. In contrast to other qubit implementations, point defect qubits in wide-bandgap semiconductors are highly coherent and robust even at elevated temperatures.\cite{weber_quantum_2010,wolfowicz_quantum_2021} Such optically addressable spin qubits, realized for example by the NV center in diamond\cite{doherty_nitrogen-vacancy_2013}, the silicon vacancy\cite{widmann_coherent_2015,ivady_identification_2017} in silicon carbide (SiC), and divacancy related defects in SiC\cite{koehl_room_2011,ivady_stabilization_2019}, can possess as long coherence time as 1~ms at room temperature.\cite{balasubramanian_ultralong_2009,christle_isolated_2015,widmann_coherent_2015} Research activities in this area have gained considerable momentum over the last decades and point defect-based quantum devices have become leading contenders in several areas of quantum technologies, such as quantum sensing and quantum internet.\cite{awschalom_quantum_2018}

The coherence of spin qubits in light element semiconductors is often limited by spin-spin interactions with paramagnetic defects and nuclei. In high-purity samples, the magnetic environment of a spin qubit is defined by the surrounding nuclear spin bath.\cite{zhao_decoherence_2012,seo_quantum_2016} Point defect spins interact with nuclear spins through the hyperfine coupling that depends on the spatial distribution of the defect's spin density and the position of the nuclear spins. The hyperfine spin Hamiltonian term is parameterized by the hyperfine tensor, whose elements can be measured by various magnetic resonance techniques and calculated by using first-principles electronic structure methods. Conventionally, electron spin resonance (ESR) has been used to determine the hyperfine tensor for the closest nuclear spins giving rise to $\sim$10-300~MHz hyperfine splitting of the nuclear spin sublevels. The high controllability and the long coherence time of point defect qubits have enabled more sophisticated nuclear spin detection techniques to be developed. Optically detected magnetic resonance (ODMR)  measurement of individual point defect qubits allowed the detection of nuclear spins 2.5-7~\AA\ distances from the NV center with hyperfine splitting ranging from 430~kHz to 14~MHz.\cite{smeltzer_13c_2011,dreau_high-resolution_2012} Dynamic decoupling techniques can be used to boost further the sensitivity of the measurements.\cite{taminiau_detection_2012,wang_positioning_2016,zopes_three-dimensional_2018,abobeih_atomic-scale_2019,van_de_stolpe_mapping_2024} Using such techniques, nuclear spins as distant as 30~\AA\ could be detected with hyperfine splitting of $\sim 1$~kHz.\cite{abobeih_atomic-scale_2019,van_de_stolpe_mapping_2024} These developments have opened new directions in magnetic resonance and magnetic resonance imaging in nanometer scales.\cite{taminiau_detection_2012,wang_positioning_2016,zopes_three-dimensional_2018,abobeih_atomic-scale_2019,van_de_stolpe_mapping_2024} In addition, the characterized nearby nuclear spins can be utilized as additional highly coherent quantum resources for quantum computation and quantum internet.\cite{bradley_ten-qubit_2019,abobeih_fault-tolerant_2022} 

Hyperfine coupling tensors can be calculated using different first principles methods, such as density functional theory (DFT) and wave functions-based methods. The accuracy of the computed hyperfine tensors for close nuclear spins is remarkable. For example, considering paramagnetic point defects in semiconductors a mean absolute relative error of 4.7\% has been reported.\cite{szasz_hyperfine_2013} Since the hyperfine structure of the point defect's spin sublevels is unique like a bar code, one can compare the measured and computed hyperfine tensors to identify paramagnetic defects in semiconductors unambiguously, see for instance Refs.~\cite{son_divacancy_2006,isoya_epr_2008,ivady_identification_2017}.

As demonstrated recently by a supercell-size-scaling test in Ref.~\cite{ivady_first_2018}, the accuracy of the computed hyperfine parameters is only limited for the closest nuclear spin in 1-5~\AA\ distances from the defect. The relative error sharply increases for nuclear spins located at large distances. As discussed in Ref.~\cite{ivady_first_2018}, this is presumably due to the periodic boundary condition and related finite size effects. 
Using finite cluster models, such as the C$_{291}$H$_{172}$ cluster and  the C$_{510}$H$_{252}$ cluster used in Refs.~\cite{nizovtsev_theoretical_2014,nizovtsev_non-flipping13c_2018} to calculate the hyperfine coupling tensors at 1.5-8~\AA\ distances from the defect, errors related to the periodic boundary condition can be eliminated. Another limitation of first principles calculations is the model size that maximizes the number of lattice sites that can be considered. 

In this work, we first demonstrate the inaccuracy of the numerical hyperfine parameters obtained with the industry standard VASP code\cite{VASP,VASP2}. To resolve the underlying methodological issues we introduce a real-space integration method and the use of a large support lattice for considering nuclear spins outside the boundaries of the supercell. To benchmark our method, we carry out large-scale calculations for the NV center in diamond using different exchange-correlation functionals and compare the numerical results with available experimental hyperfine data sets. From the comparison, we conclude that the HSE06\cite{heyd_hybrid_2003,heyd_erratum:_2006} functional with 0.2 mixing parameter performs best for the NV center in diamond resulting in a mean absolute percentage error of 1.7\% for nuclear spins 6-30~\AA\ distances from the NV center. This is a significant improvement compared to previous theoretical predictions obtained by using VASP. We show that the residual errors are likely related to the inaccurate calculation of the Fermi contact term. High accuracy hyperfine tensors of $\approx10^{4}$ lattice sites as well as volumetric hyperfine data with $<0.1$~\AA\ spatial resolution are published together with this article~\footnote{Theoretical hyperfine tensors for all lattice sites within 30~\AA\ distance from the NV center are available at \href{https://ivadygroup.elte.hu/hyperfine}{https://ivadygroup.elte.hu/hyperfine}\label{footnote1}} and ready to be used for modeling NV center quantum nodes and positioning nuclear spins around the NV center in diamond in nano-NMR measurements.

\section{Results and Discussion}
 
\begin{figure}
    \centering
        \includegraphics[width=.9\linewidth]{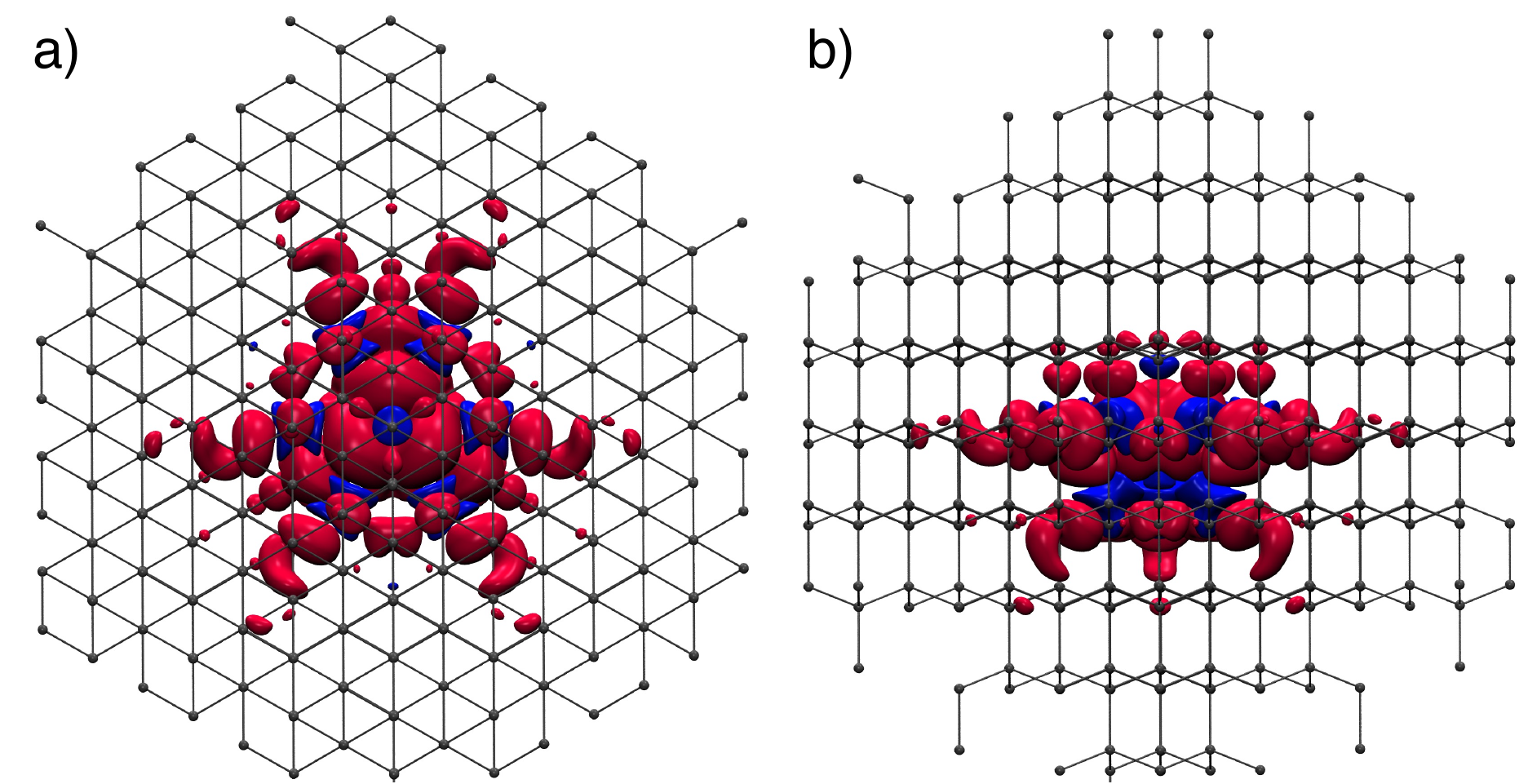}
    \caption{Spin density of the NV center in diamond. a) and b) depict top and side views, respectively. Red (blue) lobs indicate positive (negative) spin density isosurfaces and gray dots and bars show the lattice of the diamond. The isosurface value is set to $\pm$0.003. The blue lobs indicate weak antiferromagnetic couplings with neighboring atoms, e.g.\ the nitrogen of the NV center.}
    \label{fig:spin_density}
\end{figure}

The hyperfine interaction describes the weak coupling between the electron and the nuclear spins. Integrating out the spatial degrees of freedom, the corresponding hyperfine spin Hamiltonian term can be written as
\begin{equation}\label{hyp}
     H_{\text{HF}}=\mathbf{S}\mathbf{\mathcal{A}}\mathbf{I} \text{,}
\end{equation}
where $\mathbf{\mathcal{A}}$ is the hyperfine tensor and $\mathbf{S}$ and $\mathbf{I}$ are the electron spin and the nuclear spin operator vectors with $S = 1$ and $I = 1/2$ quantum numbers for the NV center and adjacent $^{13}$C nuclear spins of the diamond lattice, respectively.  The two dominant contributions to the hyperfine tensor are the Fermi contact interaction, $\mathcal{A}_{\text{FC}}$, and the magnetic dipole-dipole coupling of the electron and nuclear spins, $\mathcal{A}_{\text{SS}}$.

Elements of the hyperfine tensor can be calculated given the spin density of the electron spin $\sigma(\mathbf{r})$, see Fig.~\ref{fig:spin_density} for the NV center, and the position $\mathbf{R}_J$ of nuclear spin $J$ as \cite{szasz_hyperfine_2013}
\begin{equation}
     \mathcal{A}_{ij}^{J}=\frac{1}{2S}\gamma_{J}\gamma_{e}\hbar^{2}\Bigg[ \frac{8\pi}{3 }\int\delta(\mathbf{r}-\mathbf{R}_{J})\sigma(\mathbf{r})d\mathbf{r}+\mathcal{W}_{ij}(\mathbf{R}_{J})\Bigg]\text{ ,}
     \label{HF0}
\end{equation}
where $\gamma_{J}$ and $\gamma_{e}$ are the gyromagnetic ratio of nucleus $J$ and the electron, respectively. The first term on the r.h.s.\ of Eq.~(\ref{HF0}) accounts for the Fermi contact interactions, where the Dirac delta function $\delta (\mathbf{r}-\mathbf{R}_J)$ takes the value of the spin density at the position of the nucleus whose spatial distribution is neglected. The largest contribution to the Fermi contact term originates from electronic states of $s$ orbit character exhibiting non-zero probability at the nucleus site. The second term on the r.h.s.\ of Eq.~(\ref{HF0}) accounts for the the dipole-dipole interaction, where the integral $\mathcal{W}_{ij}$ can be expressed as
\begin{equation}\label{FC}
     \mathcal{W}_{ij}(\mathbf{R})=\int\Bigg(\frac{3(\mathbf{r}-\mathbf{R})_{i}(\mathbf{r} \mathbf{R})_{j}}{|\mathbf{r}-\mathbf{R}|^{5}}-\frac{\delta_{ij}}{|\mathbf{r}-\mathbf {R}|^{3}} \Bigg)\sigma(\mathbf{r})d\mathbf{r}\text{.}
\end{equation}

First principles electronic structure codes for solid state physics often use periodic boundary conditions, plane wave basis sets, and pseudopotentials to describe valance states of periodic lattices. For example, the industry standard VASP software package \cite{VASP,VASP2} calculates the hyperfine tensor using the method of P. E. Blöchl \cite{blochl_first-principles_2000} while taking core polarization contributions into account \cite{szasz_hyperfine_2013} in periodic boundary conditions. Employing the projector augmented wave (PAW) method \cite{PAW}, the total spin density of the system is composed of three parts\cite{blochl_first-principles_2000,szasz_hyperfine_2013}
\begin{equation}
     \sigma = \tilde{\sigma}+\sigma^{1}-\tilde{\sigma}^{1}\text{,}
     \label{HF1}
\end{equation}
where $\sigma^{1}$ and $\tilde{\sigma}^{1}$ are the atomic core-centered true and pseudo-spin densities, respectively, and $\tilde{\sigma}$ is the total spin density of the valance electrons calculated using pseudo-potentials. With this differentiation, the Fermi contact interaction can be written as 
\begin{align}
         \frac{8\pi}{3}\int\delta(\mathbf{r}-\mathbf{R_{J}})\sigma(\mathbf{r})d\mathbf{r} = 
         \frac{ 8\pi}{3}\Bigg[\sum_{\mathbf{G}}\tilde{\sigma}(\mathbf{G})e^{i\mathbf{G}\mathbf{R}}+ \\
         +\int\delta_{T}(r)\sigma^{1}_{s\mathbf{R}}(r)dr-\int\delta_{T}(r)\tilde{\sigma}^{ 1}_{s\mathbf{R}}(r)dr \Bigg]\text{,}
\end{align}
where $\sigma^{1}_{s\mathbf{R}}$ and $\tilde{\sigma}^{ 1}_{s\mathbf{R}}$ are the $s$-like contributions to the true and pseudo-core-centered spin densities respectively, and $\delta_{T}(r)$ is an extended Dirac-delta function, that takes into account the relativistic effects\cite{szasz_hyperfine_2013}. Spin polarization of the core electrons can be calculated within the frozen valence approximation.\cite{yazyev_core_2005,szasz_hyperfine_2013} The computed core polarization is added to $\sigma^{1}$ and the corresponding hyperfine contribution $\mathcal{A}_{\text{1c}}$ is determined. 
With a similar line of thought, the dipole-dipole interaction can also be expressed as 
\begin{equation} \label{eq:wij_paw}
     \mathcal{W}_{ij}(\mathbf{R})= \tilde{\mathcal{W}}_{ij}(\mathbf{R})+ \mathcal{W}_{ij}^{1 }(\mathbf{R})- \tilde{\mathcal{W}}_{ij}^{1}(\mathbf{R})\text{,}
\end{equation}
where the valance electrons' contribution $\tilde{\mathcal{W}}_{ij}$ is obtained from the pseudo-spin density $\tilde{\sigma}$ as
\begin{equation} \label{eq:wij_pseudo}
     \tilde{\mathcal{W}}_{ij}(\mathbf{R})=-4\pi\sum_{\mathbf{G}}\Bigg(\frac{G_{i}G_{j}}{ G^{2}}-\frac{\delta_{ij}}{3}\Bigg)\tilde{\sigma}(\mathbf{G})e^{i\mathbf{G}\mathbf{R}} \text{,}
\end{equation}
and the one-center contributions to the dipole-dipole term can be obtained from $d$-like contribution to the one-center spin density.\cite{blochl_first-principles_2000,szasz_hyperfine_2013}

\begin{figure}
    \centering
        \includegraphics[width=.65\linewidth]{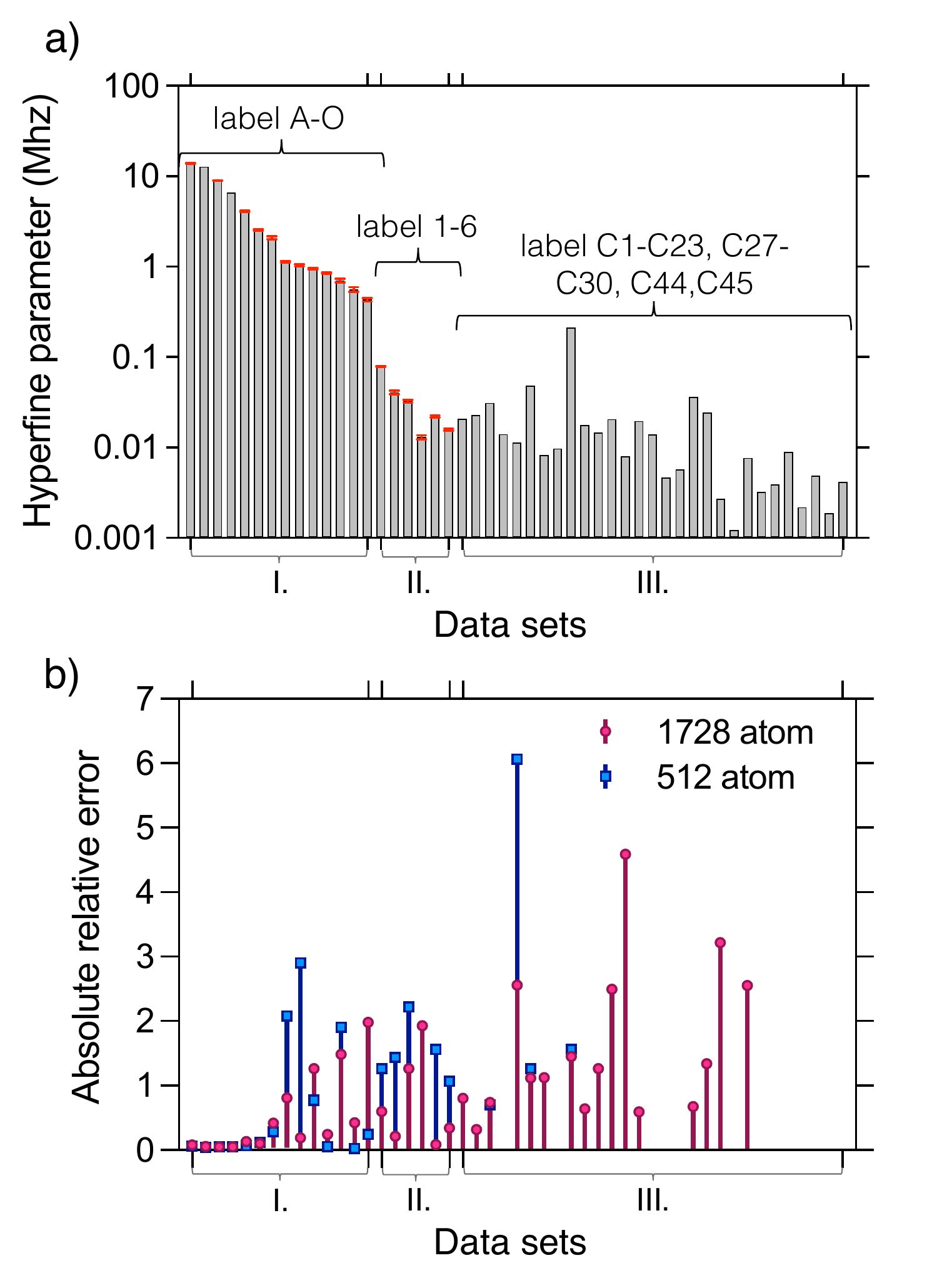}
    \caption{Comparison of experimental and theoretical hyperfine parameters. {\bf a)} Experimental data sets of hyperfine parameters reported in Ref.~\cite{dreau_high-resolution_2012} (data set I.), in Ref.~\cite{taminiau_detection_2012} (data set II.), and in Ref.~\cite{van_de_stolpe_mapping_2024} (data set III.). Gray columns depict the measured hyperfine parameters, $A_z$ for data set I.\ and  $A_{zz}$ for data sets II.\ and III. The red bars depict the absolute error of the measurements.  {\bf b)} Absolute relative error (ARE) of the calculated hyperfine values showing increasing error due to finite-size effects. Blue columns with squares and red columns with circles depict the ARE of calculations with supercells containing respectively 512 -, and 1728 atomic positions. When no bars are depicted for certain elements of the data sets, the corresponding nuclei site is outside the supercell used for the calculations. Value 1 on the $y$-scale corresponds to 100\% absolute percentage error.}
    \label{fig:orig_hyp_error}
\end{figure}

First, we use this method as implemented in VASP to calculate hyperfine tensors of all sites in a 512-atom and a 1728-atom supercell of diamond including a single NV center in the middle of the supercell, see Fig.~\ref{fig:spin_density}. For the calculations, we use HSE06 exchange-correlation functional, 500~eV cutoff energy of the plane-wave basis set, $\Gamma$-point sampling of the Brillouin zone, and high convergence criteria.  The structure of the defect is optimized as far as the largest force is smaller than 10$^{-3}$~eV/\AA . To compare our results with the experimental values, see Fig.~\ref{fig:orig_hyp_error}(a), we either compute the hyperfine splitting as $A_z = \sqrt{A_{xz}^2 + A_{yz}^2+A_{zz}^2} $ \cite{dreau_high-resolution_2012}, where $A_{ij}$ are the elements of the hyperfine tensor, which is compared with experimental values in data set I.\ taken from Ref.~\cite{dreau_high-resolution_2012}, or compare the $A_{zz}$  hyperfine tensor element with experimental $A_{zz}$ values obtained from high precision measurements in Ref.~\cite{taminiau_detection_2012} (data set II.) and Ref.~\cite{van_de_stolpe_mapping_2024} (data set III.). Note that the latter data set is a refined and extended version of the data published in Ref.~\cite{abobeih_atomic-scale_2019}. From the reported 50 nuclear spins, we consider only those for which both the position and the hyperfine parameter are reported with high accuracy, see Fig.~\ref{fig:orig_hyp_error}(a). When comparing the computed $A_{zz}$ values with data sets II.-III., we found a consistent sign difference between theory and experiment. The sign of our hyperfine values agrees with other theoretical calculations, i.e.\  VASP's values as well as the values reported in Refs.~\cite{nizovtsev_theoretical_2014,nizovtsev_non-flipping13c_2018}. Therefore, we anticipate that the discrepancy originates from the convention used in the experiments. Hereinafter, we omit this sign difference.  

In Fig.~\ref{fig:orig_hyp_error}(b), we depict the absolute relative error of the computed hyperfine splitting values for data set I.-III. As can be seen, the absolute relative error of the theoretical values rapidly increases with decreasing value of the experimental hyperfine parameter and below $\sim$1~MHz wildly fluctuates. It is also clear that the use of a large supercell does not improve the comparison with the experiments. Finally, we note that for making the comparison between theory and experiment we needed to position the nuclear spins measured in data sets I.\ and II. For this, we used our improved hyperfine tensor computation discussed next. 

Finite-size correction of hyperfine tensors computed in periodic supercell models has not been thoroughly investigated before. There are several possible sources of finite-size effects that can lead to non-physical interaction of nuclear spins and period defect structures. For instance, the spin density of a defect and its periodic replicas may overlap in small supercell models giving rise to overestimated Fermi contact interaction terms and perturbed dipole-dipole interaction terms for certain nuclei sites. This error is, however, assumed to decay exponentially with the size of the supercell since localized defect states decay exponentially. More difficult-to-handle finite-size effects arise from the long-range dipole-dipole interaction that decays with the third power of the distance of the spins. In periodic supercell models, nuclear spins interact with a lattice of defects. In contrast to the Coulomb interaction, the dipole-dipole interaction does not diverge, although the interaction strength and the hyperfine tensor's principal axis may be considerably perturbed. A nuclear spin halfway between the defect and one of its periodic replicas interacts with two spin densities with approximately the same coupling strength that can give rise to $\mathcal{O}(100\%)$ error explaining our observations depicted in Fig.~\ref{fig:orig_hyp_error}(b). In the following, we remedy these finite-size effects.

The errors are derived from the long-ranged interaction term, i.e.\  the pseudo-dipole-dipole integral $\tilde{\mathcal{W}}_{ij}$ defined in Eq.~(\ref{eq:wij_pseudo}). The periodicity of the spin density and thus the finite size effects are encoded into the pseudo spin density $\tilde{\sigma}(\mathbf{G})$. The point defect and its replicas cannot be separated in Fourier space; however, in real space, this can be easily achieved by simply limiting the range of integration. To overcome the finite-size dependence of the dipolar hyperfine interaction term, we utilize this strategy and combine it with the PAW method. To this end, we calculate  $\tilde{\mathcal{W}}_{ij}$ as
\begin{equation}\label{eq:wij_correct}
     \tilde{\mathcal{W}}_{ij}(\mathbf{R^+}) = \! \! \! \bigintss\displaylimits^{\overline{\Omega}_{\text{SC}} } \! \! \! \! \Bigg(\frac{3(\mathbf{r}-\mathbf{R^+})_{i}(\mathbf{r} - \mathbf{R^+})_{j}}{|\mathbf{r}-\mathbf{R^+}|^{5}}-\frac{\delta_{ij}}{|\mathbf{r}-\mathbf {R^+}|^{3}} \Bigg)\sigma(\mathbf{r})d\mathbf{r}\text{,}
\end{equation}
where the extended position coordinates $\mathbf{R^+}$ include the lattice sites within the supercell ($\mathbf{R}$) and outside the supercell provided by a support lattice within a sphere of 30~\AA\ radius centered on the NV center. The support lattice is aligned with the supercell, although it does not contain the atomic sites of the supercell.  This way the hyperfine interaction calculation is not limited by the size of the supercell. Note that for the integration we use the full spin density $\sigma(\mathbf{r})$ expressed on a fine grid and not the pseudo spin density in contrast to Eq.~(\ref{eq:wij_pseudo}). For nuclear spins contained within the supercell, the real-space integration is carried out over the supercell except a sphere of $r_{\text{PAW}}$ radius centered at the nuclear spin position $\mathbf{R}$, i.e.\  $\overline{\Omega}_{\text{SC}} = \Omega_{\text{SC}} - \Omega_{\text{PAW}}(\mathbf{R}) $. The total dipole-dipole interaction term is defined as $\mathcal{W}_{ij}(\mathbf{R})= \tilde{\mathcal{W}}_{ij}(\mathbf{R}) + \mathcal{W}_{ij}^{1 }(\mathbf{R})$ in contrast to Eq.~(\ref{eq:wij_paw}). The Fermi contact interaction with a core polarization contribution is obtained using the  VASP's implementation\cite{szasz_hyperfine_2013}. For lattice sites outside the supercell's boundaries, the spin density is considered to be zero at the nuclear spin site, i.e.\ both the $\mathcal{W}_{ij}^{1 }(\mathbf{R^+})$ and the Fermi contact terms are approximated to be zero. The sole non-zero term for these sites is given by Eq.~(\ref{eq:wij_correct}), where the integration is carried over the full supercell volume. With these modifications, the computed tensors account for the case when the nuclear spins interact with an isolated defect and not with a lattice of defects. We anticipate that the leading finite-size effects will be removed within our approach.

To compute the hyperfine tensors for a large number of lattice sites, $\sim$20000 sites within a sphere $R^+_{\text{cut}} = 30$~\AA\  around the NV center, we implement the real space integration method in an in-house code that post-processes the VASP output files. The ground state calculations of the NV center are carried out in 512-atom and 1728-atom supercells using the experimental lattice parameter of 3.567~\AA . We use both the semi-local PBE \cite{PBE} and various forms of the HSE06 hybrid functional, which are labeled as HSE($\alpha$), where $\alpha$ is the mixing parameters, e.g.\  HSE(0.25) = HSE06. The ground state spin density used in the calculation of Eq.~(\ref{eq:wij_correct}) is defined on a real space grid of 0.036~\AA\  spacing. The positions of the nuclear spins for data set III.\  are taken from  Ref.~\cite{van_de_stolpe_mapping_2024}. For data sets I.\ and II.\  we use the following strategy for positioning the nuclear spins: Considering an experimental hyperfine value, we look for the closest theoretical value in our data set. Depending on the error bar of the measurement and the calculations, we can position nuclear spin up to symmetrically equivalent sites with this method. 

\begin{figure}
    \centering
        \includegraphics[width=.7\linewidth]{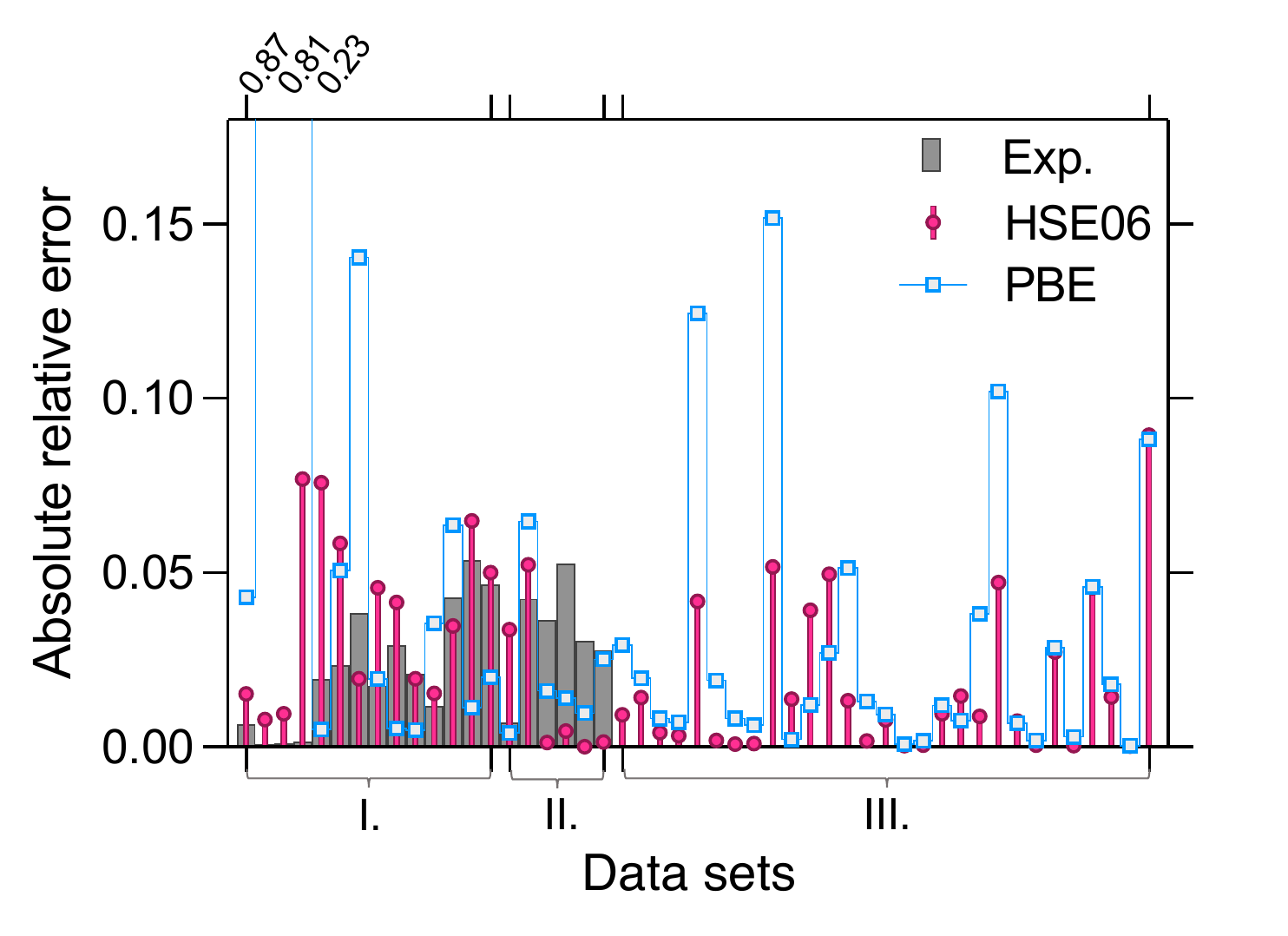}
    \caption{Absolute relative error of the hyperfine splitting values calculated with our improved integration method. Gray columns depict the absolute relative error (ARE) of the experimental data, red thin columns with circles depict the ARE of the computed hyperfine values obtained with HSE06 functional, and light blue line with squares depict the ARE of the theoretical values obtained with PBE functional, see Methods for details of the calculations. The values provided on the upper horizontal axis are the PBE absolute relative errors being out of the range of the vertical axis. }
    \label{fig:improved_hyp_error}
\end{figure}

The hyperfine splitting values computed with our method are significantly improved compared to the currently available implementation\cite{szasz_hyperfine_2013}. The relative error of the theoretical values reduce from  $\mathcal{O}(100\%)$ to $\mathcal{O}(1\%)$, see Fig.~\ref{fig:orig_hyp_error}(b) and Fig.~\ref{fig:improved_hyp_error}. For different data sets, we obtain different mean absolute percentage errors (MAPE). For data set I.\ we obtain a MAPE of 3.8\%, while the experimental data exhibit an averaged relative error margin of 1.2\%\cite{dreau_high-resolution_2012}. The obtained MAPE of the improved theoretical values for data set II.\  is  1.5\%, which is better than the relative error margin of the experimental data of 3.3\%\cite{taminiau_detection_2012}. This suggests that the theoretical values are "overfitted" for data set II., i.e.\ multiple matching hyperfine values were found within the error margin of the experimental data. By selecting the closest ones, we could obtain a MAPE smaller than the experimental error bar. The most accurate hyperfine values, with a relative error margin of $6 \times 10^{-3}\%$, together with nuclear spin positions are provided for data set III.\cite{van_de_stolpe_mapping_2024}. These high-precision measurements allow us to make a reliable assessment of the error bar of the theoretical values. Considering the 29 most accurately measured and positioned nuclear spins, we obtain a MAPE of 1.79\%. In Fig.~(\ref{fig:improved_hyp_error}), we also depicted the ARE of hyperfine values obtained with the PBE functional. As can be seen, the HSE06 functional consistently improves on the PBE values, especially for nuclear spins found close to the NV center. We also note that the use of accurate DFT spin density has high relevance. Considering point spin density approximation, i.e.\ $\sigma ( \mathbf{r}) =  \delta( \mathbf{r} - \mathbf{r}_0)$ where $\mathbf{r}_0$ is the center of the NV center, we obtain a mean absolute relative error (MARE) of 76\%  for data set III.\ see Supplementary Figure~S1.

\begin{figure}[!h]
    \centering
        \includegraphics[width=.7\linewidth]{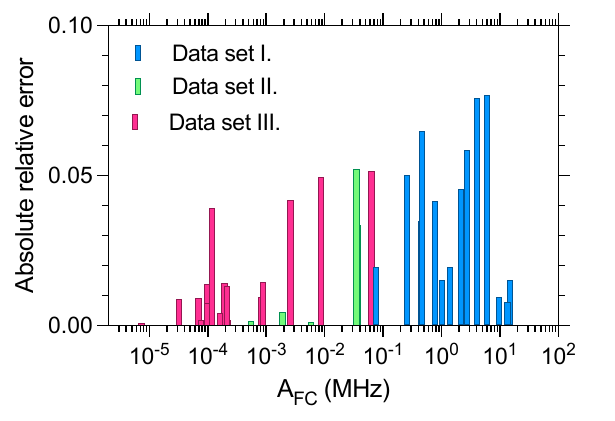}
    \caption{ Absolute relative error of the theoretical values as a function of the Fermi contact contribution to the computed hyperfine parameter. The theoretical values are obtained by using HSE06 functional and 1728 atom supercell. Green, blue, and red columns depict the absolute relative error (ARE) values corresponding to data sets I., II., and III., respectively.   }
    \label{fig:Afc-dep}
\end{figure}

Finally, we investigate the source of residual errors and possible further measures to improve the computed hyperfine tensors. For this study, we use data set III. First of all, it is notable that the mean signed relative error (MSRE) of -0.12\% is significantly smaller than the MARE of 1.79\%, suggesting that there are no large systematic errors and the discrepancies may be related to numeral uncertainties.  Next, we plot the MARE as a function of the Fermi contact term, see Fig.~(\ref{fig:Afc-dep}), and as a function of the distance, see Supplementary Figure~S2. There seems to be a correlation between the largest errors and the value of the Fermi contact term. Due to the distance dependence of the Fermi contact term,  the MARE seems to decay with the distance of the nuclear spins, see Supplementary Figure ~S2. It should be noted; however, that large errors are also obtained for a few nuclear spins that are beyond the boundaries of the supercell, where we explicitly neglect the Fermi contact terms. Here, the neglect of the Fermi contact term may be the source of the error. To attempt reducing the errors, we tune the mixing parameter $\alpha$ of the HSE functional and study the variation of the MARE obtained for data set III., see Supplementary Figure~S3. By reducing the mixing parameter from 0.25 to 0.2, we obtain a slightly decreased MARE of 1.69\%, although an enlarged MSRE of -0.32\%. These results indicate that tuning the functional's inner parameters may lead to an improved description of the spin density, which in turn can reduce the relative error between theory and experiments. Overall, the hyperfine values may be further improved by enhancing numerical accuracy for the core/Fermi contact contribution and by fine-tuning the functional.


In summary, we demonstrated in this article that high-accuracy, finite-size effect-free hyperfine tensors can be calculated using an improved integration method. Compared to the industry standard VASP code, we could achieve a $\sim$100 fold reduction of the MARE of the theoretical values. Having more experimental data of high precision and further improved numerical accuracy in larger supercells may help to obtain superior theoretical hyperfine data compared to what has been presented here. The obtained and potentially improved future data is available online~[34]. The provided data can be used for high-precision simulation of the NV center-nuclear spin few-body quantum systems as well as positioning nuclear spin around the NV center.

\section*{Methods}

For VASP calculations, we use a 512-atom and a 1728-atom supercell of diamond including a single NV center in the middle of the supercell. For the calculations, we use the Perdew–Burke–Ernzerhof (PBE)\cite{PBE} and the Heyd–Scuseria–Ernzerhof (HSE06)\cite{heyd_hybrid_2003,heyd_erratum:_2006} exchange-correlation functional, 500~eV cutoff energy of the plane-wave basis set, $\Gamma$-point sampling of the Brillouin zone, and high convergence criteria (PREC = Accurate).  The energy threshold for the self-consistent field calculations is set to 10$^{-6}$~eV. The structure of the defect is optimized as far as the largest force is smaller than 10$^{-3}$~eV/\AA . For the real-space hyperfine tensor calculations, we use the convergent spin density of the 1728-atom supercell model defined on a fine real-space grid of 0.036~\AA\ ($a_0$/600) spacing obtained by VASP.

\section*{Data availability}

The calculated hyperfine tensors for all lattice sites within 30~\AA\  distance from the NV center are available at \href{https://ivadygroup.elte.hu/hyperfine}{https://ivadygroup.elte.hu/hyperfine}.

\section*{Code availability}

The real-space integration code that postprocesses VASP outputs is available at \newline
\href{https://ivadygroup.elte.hu/hyperfine}{https://ivadygroup.elte.hu/hyperfine}.


\begin{thebibliography}{38}%
\makeatletter
\providecommand \@ifxundefined [1]{%
 \@ifx{#1\undefined}
}%
\providecommand \@ifnum [1]{%
 \ifnum #1\expandafter \@firstoftwo
 \else \expandafter \@secondoftwo
 \fi
}%
\providecommand \@ifx [1]{%
 \ifx #1\expandafter \@firstoftwo
 \else \expandafter \@secondoftwo
 \fi
}%
\providecommand \natexlab [1]{#1}%
\providecommand \enquote  [1]{``#1''}%
\providecommand \bibnamefont  [1]{#1}%
\providecommand \bibfnamefont [1]{#1}%
\providecommand \citenamefont [1]{#1}%
\providecommand \href@noop [0]{\@secondoftwo}%
\providecommand \href [0]{\begingroup \@sanitize@url \@href}%
\providecommand \@href[1]{\@@startlink{#1}\@@href}%
\providecommand \@@href[1]{\endgroup#1\@@endlink}%
\providecommand \@sanitize@url [0]{\catcode `\\12\catcode `\$12\catcode
  `\&12\catcode `\#12\catcode `\^12\catcode `\_12\catcode `\%12\relax}%
\providecommand \@@startlink[1]{}%
\providecommand \@@endlink[0]{}%
\providecommand \url  [0]{\begingroup\@sanitize@url \@url }%
\providecommand \@url [1]{\endgroup\@href {#1}{\urlprefix }}%
\providecommand \urlprefix  [0]{URL }%
\providecommand \Eprint [0]{\href }%
\providecommand \doibase [0]{http://dx.doi.org/}%
\providecommand \selectlanguage [0]{\@gobble}%
\providecommand \bibinfo  [0]{\@secondoftwo}%
\providecommand \bibfield  [0]{\@secondoftwo}%
\providecommand \translation [1]{[#1]}%
\providecommand \BibitemOpen [0]{}%
\providecommand \bibitemStop [0]{}%
\providecommand \bibitemNoStop [0]{.\EOS\space}%
\providecommand \EOS [0]{\spacefactor3000\relax}%
\providecommand \BibitemShut  [1]{\csname bibitem#1\endcsname}%
\let\auto@bib@innerbib\@empty
\bibitem [{\citenamefont {Jelezko}\ \emph {et~al.}(2001)\citenamefont
  {Jelezko}, \citenamefont {Tietz}, \citenamefont {Gruber}, \citenamefont
  {Popa}, \citenamefont {Nizovtsev}, \citenamefont {Kilin},\ and\ \citenamefont
  {Wrachtrup}}]{jelezko_spectroscopy_2001}%
  \BibitemOpen
  \bibfield  {author} {\bibinfo {author} {\bibfnamefont {F.}~\bibnamefont
  {Jelezko}}, \bibinfo {author} {\bibfnamefont {C.}~\bibnamefont {Tietz}},
  \bibinfo {author} {\bibfnamefont {A.}~\bibnamefont {Gruber}}, \bibinfo
  {author} {\bibfnamefont {I.}~\bibnamefont {Popa}}, \bibinfo {author}
  {\bibfnamefont {A.}~\bibnamefont {Nizovtsev}}, \bibinfo {author}
  {\bibfnamefont {S.}~\bibnamefont {Kilin}}, \ and\ \bibinfo {author}
  {\bibfnamefont {J.}~\bibnamefont {Wrachtrup}},\ }\bibfield  {title} {\enquote
  {\bibinfo {title} {Spectroscopy of single n-v centers in diamond},}\
  }\href@noop {} {\bibfield  {journal} {\bibinfo  {journal} {Single Molecules}\
  }\textbf {\bibinfo {volume} {2}},\ \bibinfo {pages} {255--260} (\bibinfo
  {year} {2001})}\BibitemShut {NoStop}%
\bibitem [{\citenamefont {Weber}\ \emph {et~al.}(2010)\citenamefont {Weber},
  \citenamefont {Koehl}, \citenamefont {Varley}, \citenamefont {Janotti},
  \citenamefont {Buckley}, \citenamefont {Van~de Walle},\ and\ \citenamefont
  {Awschalom}}]{weber_quantum_2010}%
  \BibitemOpen
  \bibfield  {author} {\bibinfo {author} {\bibfnamefont {J.~R.}\ \bibnamefont
  {Weber}}, \bibinfo {author} {\bibfnamefont {W.~F.}\ \bibnamefont {Koehl}},
  \bibinfo {author} {\bibfnamefont {J.~B.}\ \bibnamefont {Varley}}, \bibinfo
  {author} {\bibfnamefont {A.}~\bibnamefont {Janotti}}, \bibinfo {author}
  {\bibfnamefont {B.~B.}\ \bibnamefont {Buckley}}, \bibinfo {author}
  {\bibfnamefont {C.~G.}\ \bibnamefont {Van~de Walle}}, \ and\ \bibinfo
  {author} {\bibfnamefont {D.~D.}\ \bibnamefont {Awschalom}},\ }\bibfield
  {title} {\enquote {\bibinfo {title} {Quantum computing with defects},}\
  }\href {\doibase 10.1073/pnas.1003052107} {\bibfield  {journal} {\bibinfo
  {journal} {Proc. Natl. Acad. Sci.}\ }\textbf {\bibinfo {volume} {107}},\
  \bibinfo {pages} {8513--8518} (\bibinfo {year} {2010})}\BibitemShut {NoStop}%
\bibitem [{\citenamefont {Wehner}\ \emph {et~al.}(2018)\citenamefont {Wehner},
  \citenamefont {Elkouss},\ and\ \citenamefont {Hanson}}]{wehner_quantum_2018}%
  \BibitemOpen
  \bibfield  {author} {\bibinfo {author} {\bibfnamefont {Stephanie}\
  \bibnamefont {Wehner}}, \bibinfo {author} {\bibfnamefont {David}\
  \bibnamefont {Elkouss}}, \ and\ \bibinfo {author} {\bibfnamefont {Ronald}\
  \bibnamefont {Hanson}},\ }\bibfield  {title} {\enquote {\bibinfo {title}
  {Quantum internet: A vision for the road ahead},}\ }\href {\doibase
  10.1126/science.aam9288} {\bibfield  {journal} {\bibinfo  {journal}
  {Science}\ }\textbf {\bibinfo {volume} {362}} (\bibinfo {year} {2018}),\
  10.1126/science.aam9288}\BibitemShut {NoStop}%
\bibitem [{\citenamefont {Bradley}\ \emph {et~al.}()\citenamefont {Bradley},
  \citenamefont {Randall}, \citenamefont {Abobeih}, \citenamefont {Berrevoets},
  \citenamefont {Degen}, \citenamefont {Bakker}, \citenamefont {Markham},
  \citenamefont {Twitchen},\ and\ \citenamefont
  {Taminiau}}]{bradley_ten-qubit_2019}%
  \BibitemOpen
  \bibfield  {author} {\bibinfo {author} {\bibfnamefont {C.E.}\ \bibnamefont
  {Bradley}}, \bibinfo {author} {\bibfnamefont {J.}~\bibnamefont {Randall}},
  \bibinfo {author} {\bibfnamefont {M.H.}\ \bibnamefont {Abobeih}}, \bibinfo
  {author} {\bibfnamefont {R.C.}\ \bibnamefont {Berrevoets}}, \bibinfo {author}
  {\bibfnamefont {M.J.}\ \bibnamefont {Degen}}, \bibinfo {author}
  {\bibfnamefont {M.A.}\ \bibnamefont {Bakker}}, \bibinfo {author}
  {\bibfnamefont {M.}~\bibnamefont {Markham}}, \bibinfo {author} {\bibfnamefont
  {D.J.}\ \bibnamefont {Twitchen}}, \ and\ \bibinfo {author} {\bibfnamefont
  {T.H.}\ \bibnamefont {Taminiau}},\ }\bibfield  {title} {\enquote {\bibinfo
  {title} {A ten-qubit solid-state spin register with quantum memory up to one
  minute},}\ }\href {\doibase 10.1103/PhysRevX.9.031045} {\bibfield  {journal}
  {\bibinfo  {journal} {Physical Review X}\ }\textbf {\bibinfo {volume} {9}},\
  \bibinfo {pages} {031045}}\BibitemShut {NoStop}%
\bibitem [{\citenamefont {Abobeih}\ \emph {et~al.}()\citenamefont {Abobeih},
  \citenamefont {Wang}, \citenamefont {Randall}, \citenamefont {Loenen},
  \citenamefont {Bradley}, \citenamefont {Markham}, \citenamefont {Twitchen},
  \citenamefont {Terhal},\ and\ \citenamefont
  {Taminiau}}]{abobeih_fault-tolerant_2022}%
  \BibitemOpen
  \bibfield  {author} {\bibinfo {author} {\bibfnamefont {M.~H.}\ \bibnamefont
  {Abobeih}}, \bibinfo {author} {\bibfnamefont {Y.}~\bibnamefont {Wang}},
  \bibinfo {author} {\bibfnamefont {J.}~\bibnamefont {Randall}}, \bibinfo
  {author} {\bibfnamefont {S.~J.~H.}\ \bibnamefont {Loenen}}, \bibinfo {author}
  {\bibfnamefont {C.~E.}\ \bibnamefont {Bradley}}, \bibinfo {author}
  {\bibfnamefont {M.}~\bibnamefont {Markham}}, \bibinfo {author} {\bibfnamefont
  {D.~J.}\ \bibnamefont {Twitchen}}, \bibinfo {author} {\bibfnamefont {B.~M.}\
  \bibnamefont {Terhal}}, \ and\ \bibinfo {author} {\bibfnamefont {T.~H.}\
  \bibnamefont {Taminiau}},\ }\bibfield  {title} {\enquote {\bibinfo {title}
  {Fault-tolerant operation of a logical qubit in a diamond quantum
  processor},}\ }\href {\doibase 10.1038/s41586-022-04819-6} {\bibfield
  {journal} {\bibinfo  {journal} {Nature}\ }\textbf {\bibinfo {volume} {606}},\
  \bibinfo {pages} {884--889}}\BibitemShut {NoStop}%
\bibitem [{\citenamefont {Wolfowicz}\ \emph {et~al.}(2021)\citenamefont
  {Wolfowicz}, \citenamefont {Heremans}, \citenamefont {Anderson},
  \citenamefont {Kanai}, \citenamefont {Seo}, \citenamefont {Gali},
  \citenamefont {Galli},\ and\ \citenamefont
  {Awschalom}}]{wolfowicz_quantum_2021}%
  \BibitemOpen
  \bibfield  {author} {\bibinfo {author} {\bibfnamefont {Gary}\ \bibnamefont
  {Wolfowicz}}, \bibinfo {author} {\bibfnamefont {F.~Joseph}\ \bibnamefont
  {Heremans}}, \bibinfo {author} {\bibfnamefont {Christopher~P.}\ \bibnamefont
  {Anderson}}, \bibinfo {author} {\bibfnamefont {Shun}\ \bibnamefont {Kanai}},
  \bibinfo {author} {\bibfnamefont {Hosung}\ \bibnamefont {Seo}}, \bibinfo
  {author} {\bibfnamefont {Adam}\ \bibnamefont {Gali}}, \bibinfo {author}
  {\bibfnamefont {Giulia}\ \bibnamefont {Galli}}, \ and\ \bibinfo {author}
  {\bibfnamefont {David~D.}\ \bibnamefont {Awschalom}},\ }\bibfield  {title}
  {\enquote {\bibinfo {title} {Quantum guidelines for solid-state spin
  defects},}\ }\href {\doibase 10.1038/s41578-021-00306-y} {\bibfield
  {journal} {\bibinfo  {journal} {Nature Reviews Materials}\ ,\ \bibinfo
  {pages} {1--20}} (\bibinfo {year} {2021})}\BibitemShut {NoStop}%
\bibitem [{\citenamefont {Doherty}\ \emph {et~al.}(2013)\citenamefont
  {Doherty}, \citenamefont {Manson}, \citenamefont {Delaney}, \citenamefont
  {Jelezko}, \citenamefont {Wrachtrup},\ and\ \citenamefont
  {Hollenberg}}]{doherty_nitrogen-vacancy_2013}%
  \BibitemOpen
  \bibfield  {author} {\bibinfo {author} {\bibfnamefont {Marcus~W.}\
  \bibnamefont {Doherty}}, \bibinfo {author} {\bibfnamefont {Neil~B.}\
  \bibnamefont {Manson}}, \bibinfo {author} {\bibfnamefont {Paul}\ \bibnamefont
  {Delaney}}, \bibinfo {author} {\bibfnamefont {Fedor}\ \bibnamefont
  {Jelezko}}, \bibinfo {author} {\bibfnamefont {Jörg}\ \bibnamefont
  {Wrachtrup}}, \ and\ \bibinfo {author} {\bibfnamefont {Lloyd C.~L.}\
  \bibnamefont {Hollenberg}},\ }\bibfield  {title} {\enquote {\bibinfo {title}
  {The nitrogen-vacancy colour centre in diamond},}\ }\href {\doibase
  10.1016/j.physrep.2013.02.001} {\bibfield  {journal} {\bibinfo  {journal}
  {Physics Reports}\ }\textbf {\bibinfo {volume} {528}},\ \bibinfo {pages}
  {1--45} (\bibinfo {year} {2013})}\BibitemShut {NoStop}%
\bibitem [{\citenamefont {Widmann}\ \emph {et~al.}(2015)\citenamefont
  {Widmann}, \citenamefont {Lee}, \citenamefont {Rendler}, \citenamefont {Son},
  \citenamefont {Fedder}, \citenamefont {Paik}, \citenamefont {Yang},
  \citenamefont {Zhao}, \citenamefont {Yang}, \citenamefont {Booker},
  \citenamefont {Denisenko}, \citenamefont {Jamali}, \citenamefont
  {Momenzadeh}, \citenamefont {Gerhardt}, \citenamefont {Ohshima},
  \citenamefont {Gali}, \citenamefont {Janzén},\ and\ \citenamefont
  {Wrachtrup}}]{widmann_coherent_2015}%
  \BibitemOpen
  \bibfield  {author} {\bibinfo {author} {\bibfnamefont {Matthias}\
  \bibnamefont {Widmann}}, \bibinfo {author} {\bibfnamefont {Sang-Yun}\
  \bibnamefont {Lee}}, \bibinfo {author} {\bibfnamefont {Torsten}\ \bibnamefont
  {Rendler}}, \bibinfo {author} {\bibfnamefont {Nguyen~Tien}\ \bibnamefont
  {Son}}, \bibinfo {author} {\bibfnamefont {Helmut}\ \bibnamefont {Fedder}},
  \bibinfo {author} {\bibfnamefont {Seoyoung}\ \bibnamefont {Paik}}, \bibinfo
  {author} {\bibfnamefont {Li-Ping}\ \bibnamefont {Yang}}, \bibinfo {author}
  {\bibfnamefont {Nan}\ \bibnamefont {Zhao}}, \bibinfo {author} {\bibfnamefont
  {Sen}\ \bibnamefont {Yang}}, \bibinfo {author} {\bibfnamefont {Ian}\
  \bibnamefont {Booker}}, \bibinfo {author} {\bibfnamefont {Andrej}\
  \bibnamefont {Denisenko}}, \bibinfo {author} {\bibfnamefont {Mohammad}\
  \bibnamefont {Jamali}}, \bibinfo {author} {\bibfnamefont {S.~Ali}\
  \bibnamefont {Momenzadeh}}, \bibinfo {author} {\bibfnamefont {Ilja}\
  \bibnamefont {Gerhardt}}, \bibinfo {author} {\bibfnamefont {Takeshi}\
  \bibnamefont {Ohshima}}, \bibinfo {author} {\bibfnamefont {Adam}\
  \bibnamefont {Gali}}, \bibinfo {author} {\bibfnamefont {Erik}\ \bibnamefont
  {Janzén}}, \ and\ \bibinfo {author} {\bibfnamefont {Jörg}\ \bibnamefont
  {Wrachtrup}},\ }\bibfield  {title} {\enquote {\bibinfo {title} {Coherent
  control of single spins in silicon carbide at room temperature},}\ }\href
  {\doibase 10.1038/nmat4145} {\bibfield  {journal} {\bibinfo  {journal}
  {Nature Materials}\ }\textbf {\bibinfo {volume} {14}},\ \bibinfo {pages}
  {164--168} (\bibinfo {year} {2015})}\BibitemShut {NoStop}%
\bibitem [{\citenamefont {Ivády}\ \emph {et~al.}(2017)\citenamefont {Ivády},
  \citenamefont {Davidsson}, \citenamefont {Son}, \citenamefont {Ohshima},
  \citenamefont {Abrikosov},\ and\ \citenamefont
  {Gali}}]{ivady_identification_2017}%
  \BibitemOpen
  \bibfield  {author} {\bibinfo {author} {\bibfnamefont {Viktor}\ \bibnamefont
  {Ivády}}, \bibinfo {author} {\bibfnamefont {Joel}\ \bibnamefont
  {Davidsson}}, \bibinfo {author} {\bibfnamefont {Nguyen~Tien}\ \bibnamefont
  {Son}}, \bibinfo {author} {\bibfnamefont {Takeshi}\ \bibnamefont {Ohshima}},
  \bibinfo {author} {\bibfnamefont {Igor~A.}\ \bibnamefont {Abrikosov}}, \ and\
  \bibinfo {author} {\bibfnamefont {Adam}\ \bibnamefont {Gali}},\ }\bibfield
  {title} {\enquote {\bibinfo {title} {Identification of si-vacancy related
  room-temperature qubits in \$4h\$ silicon carbide},}\ }\href {\doibase
  10.1103/PhysRevB.96.161114} {\bibfield  {journal} {\bibinfo  {journal}
  {Physical Review B}\ }\textbf {\bibinfo {volume} {96}},\ \bibinfo {pages}
  {161114} (\bibinfo {year} {2017})}\BibitemShut {NoStop}%
\bibitem [{\citenamefont {Koehl}\ \emph {et~al.}(2011)\citenamefont {Koehl},
  \citenamefont {Buckley}, \citenamefont {Heremans}, \citenamefont {Calusine},\
  and\ \citenamefont {Awschalom}}]{koehl_room_2011}%
  \BibitemOpen
  \bibfield  {author} {\bibinfo {author} {\bibfnamefont {William~F.}\
  \bibnamefont {Koehl}}, \bibinfo {author} {\bibfnamefont {Bob~B.}\
  \bibnamefont {Buckley}}, \bibinfo {author} {\bibfnamefont {F.~Joseph}\
  \bibnamefont {Heremans}}, \bibinfo {author} {\bibfnamefont {Greg}\
  \bibnamefont {Calusine}}, \ and\ \bibinfo {author} {\bibfnamefont {David~D.}\
  \bibnamefont {Awschalom}},\ }\bibfield  {title} {\enquote {\bibinfo {title}
  {Room temperature coherent control of defect spin qubits in silicon
  carbide},}\ }\href {\doibase 10.1038/nature10562} {\bibfield  {journal}
  {\bibinfo  {journal} {Nature}\ }\textbf {\bibinfo {volume} {479}},\ \bibinfo
  {pages} {84--87} (\bibinfo {year} {2011})}\BibitemShut {NoStop}%
\bibitem [{\citenamefont {Ivády}\ \emph {et~al.}(2019)\citenamefont {Ivády},
  \citenamefont {Davidsson}, \citenamefont {Delegan}, \citenamefont {Falk},
  \citenamefont {Klimov}, \citenamefont {Whiteley}, \citenamefont
  {Hruszkewycz}, \citenamefont {Holt}, \citenamefont {Heremans}, \citenamefont
  {Son}, \citenamefont {Awschalom}, \citenamefont {Abrikosov},\ and\
  \citenamefont {Gali}}]{ivady_stabilization_2019}%
  \BibitemOpen
  \bibfield  {author} {\bibinfo {author} {\bibfnamefont {Viktor}\ \bibnamefont
  {Ivády}}, \bibinfo {author} {\bibfnamefont {Joel}\ \bibnamefont
  {Davidsson}}, \bibinfo {author} {\bibfnamefont {Nazar}\ \bibnamefont
  {Delegan}}, \bibinfo {author} {\bibfnamefont {Abram~L.}\ \bibnamefont
  {Falk}}, \bibinfo {author} {\bibfnamefont {Paul~V.}\ \bibnamefont {Klimov}},
  \bibinfo {author} {\bibfnamefont {Samuel~J.}\ \bibnamefont {Whiteley}},
  \bibinfo {author} {\bibfnamefont {Stephan~O.}\ \bibnamefont {Hruszkewycz}},
  \bibinfo {author} {\bibfnamefont {Martin~V.}\ \bibnamefont {Holt}}, \bibinfo
  {author} {\bibfnamefont {F.~Joseph}\ \bibnamefont {Heremans}}, \bibinfo
  {author} {\bibfnamefont {Nguyen~Tien}\ \bibnamefont {Son}}, \bibinfo {author}
  {\bibfnamefont {David~D.}\ \bibnamefont {Awschalom}}, \bibinfo {author}
  {\bibfnamefont {Igor~A.}\ \bibnamefont {Abrikosov}}, \ and\ \bibinfo {author}
  {\bibfnamefont {Adam}\ \bibnamefont {Gali}},\ }\bibfield  {title} {\enquote
  {\bibinfo {title} {Stabilization of point-defect spin qubits by quantum
  wells},}\ }\href {\doibase 10.1038/s41467-019-13495-6} {\bibfield  {journal}
  {\bibinfo  {journal} {Nature Communications}\ }\textbf {\bibinfo {volume}
  {10}},\ \bibinfo {pages} {5607} (\bibinfo {year} {2019})}\BibitemShut
  {NoStop}%
\bibitem [{\citenamefont {Balasubramanian}\ \emph {et~al.}(2009)\citenamefont
  {Balasubramanian}, \citenamefont {Neumann}, \citenamefont {Twitchen},
  \citenamefont {Markham}, \citenamefont {Kolesov}, \citenamefont {Mizuochi},
  \citenamefont {Isoya}, \citenamefont {Achard}, \citenamefont {Beck},
  \citenamefont {Tissler}, \citenamefont {Jacques}, \citenamefont {Hemmer},
  \citenamefont {Jelezko},\ and\ \citenamefont
  {Wrachtrup}}]{balasubramanian_ultralong_2009}%
  \BibitemOpen
  \bibfield  {author} {\bibinfo {author} {\bibfnamefont {Gopalakrishnan}\
  \bibnamefont {Balasubramanian}}, \bibinfo {author} {\bibfnamefont {Philipp}\
  \bibnamefont {Neumann}}, \bibinfo {author} {\bibfnamefont {Daniel}\
  \bibnamefont {Twitchen}}, \bibinfo {author} {\bibfnamefont {Matthew}\
  \bibnamefont {Markham}}, \bibinfo {author} {\bibfnamefont {Roman}\
  \bibnamefont {Kolesov}}, \bibinfo {author} {\bibfnamefont {Norikazu}\
  \bibnamefont {Mizuochi}}, \bibinfo {author} {\bibfnamefont {Junichi}\
  \bibnamefont {Isoya}}, \bibinfo {author} {\bibfnamefont {Jocelyn}\
  \bibnamefont {Achard}}, \bibinfo {author} {\bibfnamefont {Johannes}\
  \bibnamefont {Beck}}, \bibinfo {author} {\bibfnamefont {Julia}\ \bibnamefont
  {Tissler}}, \bibinfo {author} {\bibfnamefont {Vincent}\ \bibnamefont
  {Jacques}}, \bibinfo {author} {\bibfnamefont {Philip~R.}\ \bibnamefont
  {Hemmer}}, \bibinfo {author} {\bibfnamefont {Fedor}\ \bibnamefont {Jelezko}},
  \ and\ \bibinfo {author} {\bibfnamefont {Jörg}\ \bibnamefont {Wrachtrup}},\
  }\bibfield  {title} {\enquote {\bibinfo {title} {Ultralong spin coherence
  time in isotopically engineered diamond},}\ }\href {\doibase
  10.1038/nmat2420} {\bibfield  {journal} {\bibinfo  {journal} {Nature
  Materials}\ }\textbf {\bibinfo {volume} {8}},\ \bibinfo {pages} {383--387}
  (\bibinfo {year} {2009})}\BibitemShut {NoStop}%
\bibitem [{\citenamefont {Christle}\ \emph {et~al.}(2015)\citenamefont
  {Christle}, \citenamefont {Falk}, \citenamefont {Andrich}, \citenamefont
  {Klimov}, \citenamefont {Hassan}, \citenamefont {Son}, \citenamefont
  {Janzén}, \citenamefont {Ohshima},\ and\ \citenamefont
  {Awschalom}}]{christle_isolated_2015}%
  \BibitemOpen
  \bibfield  {author} {\bibinfo {author} {\bibfnamefont {David~J.}\
  \bibnamefont {Christle}}, \bibinfo {author} {\bibfnamefont {Abram~L.}\
  \bibnamefont {Falk}}, \bibinfo {author} {\bibfnamefont {Paolo}\ \bibnamefont
  {Andrich}}, \bibinfo {author} {\bibfnamefont {Paul~V.}\ \bibnamefont
  {Klimov}}, \bibinfo {author} {\bibfnamefont {Jawad~Ul}\ \bibnamefont
  {Hassan}}, \bibinfo {author} {\bibfnamefont {Nguyen~T.}\ \bibnamefont {Son}},
  \bibinfo {author} {\bibfnamefont {Erik}\ \bibnamefont {Janzén}}, \bibinfo
  {author} {\bibfnamefont {Takeshi}\ \bibnamefont {Ohshima}}, \ and\ \bibinfo
  {author} {\bibfnamefont {David~D.}\ \bibnamefont {Awschalom}},\ }\bibfield
  {title} {\enquote {\bibinfo {title} {Isolated electron spins in silicon
  carbide with millisecond coherence times},}\ }\href {\doibase
  10.1038/nmat4144} {\bibfield  {journal} {\bibinfo  {journal} {Nature
  Materials}\ }\textbf {\bibinfo {volume} {14}},\ \bibinfo {pages} {160--163}
  (\bibinfo {year} {2015})}\BibitemShut {NoStop}%
\bibitem [{\citenamefont {Awschalom}\ \emph {et~al.}(2018)\citenamefont
  {Awschalom}, \citenamefont {Hanson}, \citenamefont {Wrachtrup},\ and\
  \citenamefont {Zhou}}]{awschalom_quantum_2018}%
  \BibitemOpen
  \bibfield  {author} {\bibinfo {author} {\bibfnamefont {David~D.}\
  \bibnamefont {Awschalom}}, \bibinfo {author} {\bibfnamefont {Ronald}\
  \bibnamefont {Hanson}}, \bibinfo {author} {\bibfnamefont {Jörg}\
  \bibnamefont {Wrachtrup}}, \ and\ \bibinfo {author} {\bibfnamefont
  {Brian~B.}\ \bibnamefont {Zhou}},\ }\bibfield  {title} {\enquote {\bibinfo
  {title} {Quantum technologies with optically interfaced solid-state spins},}\
  }\href {\doibase 10.1038/s41566-018-0232-2} {\bibfield  {journal} {\bibinfo
  {journal} {Nature Photonics}\ }\textbf {\bibinfo {volume} {12}},\ \bibinfo
  {pages} {516--527} (\bibinfo {year} {2018})}\BibitemShut {NoStop}%
\bibitem [{\citenamefont {Zhao}\ \emph {et~al.}(2012)\citenamefont {Zhao},
  \citenamefont {Ho},\ and\ \citenamefont {Liu}}]{zhao_decoherence_2012}%
  \BibitemOpen
  \bibfield  {author} {\bibinfo {author} {\bibfnamefont {Nan}\ \bibnamefont
  {Zhao}}, \bibinfo {author} {\bibfnamefont {Sai-Wah}\ \bibnamefont {Ho}}, \
  and\ \bibinfo {author} {\bibfnamefont {Ren-Bao}\ \bibnamefont {Liu}},\
  }\bibfield  {title} {\enquote {\bibinfo {title} {Decoherence and dynamical
  decoupling control of nitrogen vacancy center electron spins in nuclear spin
  baths},}\ }\href {\doibase 10.1103/PhysRevB.85.115303} {\bibfield  {journal}
  {\bibinfo  {journal} {Physical Review B}\ }\textbf {\bibinfo {volume} {85}},\
  \bibinfo {pages} {115303} (\bibinfo {year} {2012})}\BibitemShut {NoStop}%
\bibitem [{\citenamefont {Seo}\ \emph {et~al.}(2016)\citenamefont {Seo},
  \citenamefont {Falk}, \citenamefont {Klimov}, \citenamefont {Miao},
  \citenamefont {Galli},\ and\ \citenamefont {Awschalom}}]{seo_quantum_2016}%
  \BibitemOpen
  \bibfield  {author} {\bibinfo {author} {\bibfnamefont {Hosung}\ \bibnamefont
  {Seo}}, \bibinfo {author} {\bibfnamefont {Abram~L.}\ \bibnamefont {Falk}},
  \bibinfo {author} {\bibfnamefont {Paul~V.}\ \bibnamefont {Klimov}}, \bibinfo
  {author} {\bibfnamefont {Kevin~C.}\ \bibnamefont {Miao}}, \bibinfo {author}
  {\bibfnamefont {Giulia}\ \bibnamefont {Galli}}, \ and\ \bibinfo {author}
  {\bibfnamefont {David~D.}\ \bibnamefont {Awschalom}},\ }\bibfield  {title}
  {\enquote {\bibinfo {title} {Quantum decoherence dynamics of divacancy spins
  in silicon carbide},}\ }\href {\doibase 10.1038/ncomms12935} {\bibfield
  {journal} {\bibinfo  {journal} {Nature Communications}\ }\textbf {\bibinfo
  {volume} {7}},\ \bibinfo {pages} {12935} (\bibinfo {year}
  {2016})}\BibitemShut {NoStop}%
\bibitem [{\citenamefont {Smeltzer}\ \emph {et~al.}()\citenamefont {Smeltzer},
  \citenamefont {Childress},\ and\ \citenamefont {Gali}}]{smeltzer_13c_2011}%
  \BibitemOpen
  \bibfield  {author} {\bibinfo {author} {\bibfnamefont {Benjamin}\
  \bibnamefont {Smeltzer}}, \bibinfo {author} {\bibfnamefont {Lilian}\
  \bibnamefont {Childress}}, \ and\ \bibinfo {author} {\bibfnamefont {Adam}\
  \bibnamefont {Gali}},\ }\bibfield  {title} {\enquote {\bibinfo {title} {13c
  hyperfine interactions in the nitrogen-vacancy centre in diamond},}\ }\href
  {\doibase 10.1088/1367-2630/13/2/025021} {\bibfield  {journal} {\bibinfo
  {journal} {New Journal of Physics}\ }\textbf {\bibinfo {volume} {13}},\
  \bibinfo {pages} {025021}}\BibitemShut {NoStop}%
\bibitem [{\citenamefont {Dréau}\ \emph {et~al.}(2012)\citenamefont {Dréau},
  \citenamefont {Maze}, \citenamefont {Lesik}, \citenamefont {Roch},\ and\
  \citenamefont {Jacques}}]{dreau_high-resolution_2012}%
  \BibitemOpen
  \bibfield  {author} {\bibinfo {author} {\bibfnamefont {A.}~\bibnamefont
  {Dréau}}, \bibinfo {author} {\bibfnamefont {J.-R.}\ \bibnamefont {Maze}},
  \bibinfo {author} {\bibfnamefont {M.}~\bibnamefont {Lesik}}, \bibinfo
  {author} {\bibfnamefont {J.-F.}\ \bibnamefont {Roch}}, \ and\ \bibinfo
  {author} {\bibfnamefont {V.}~\bibnamefont {Jacques}},\ }\bibfield  {title}
  {\enquote {\bibinfo {title} {High-resolution spectroscopy of single {NV}
  defects coupled with nearby 13 c nuclear spins in diamond},}\ }\href
  {\doibase 10.1103/PhysRevB.85.134107} {\bibfield  {journal} {\bibinfo
  {journal} {Physical Review B}\ }\textbf {\bibinfo {volume} {85}},\ \bibinfo
  {pages} {134107} (\bibinfo {year} {2012})}\BibitemShut {NoStop}%
\bibitem [{\citenamefont {Taminiau}\ \emph {et~al.}(2012)\citenamefont
  {Taminiau}, \citenamefont {Wagenaar}, \citenamefont {van~der Sar},
  \citenamefont {Jelezko}, \citenamefont {Dobrovitski},\ and\ \citenamefont
  {Hanson}}]{taminiau_detection_2012}%
  \BibitemOpen
  \bibfield  {author} {\bibinfo {author} {\bibfnamefont {T.~H.}\ \bibnamefont
  {Taminiau}}, \bibinfo {author} {\bibfnamefont {J.~J.~T.}\ \bibnamefont
  {Wagenaar}}, \bibinfo {author} {\bibfnamefont {T.}~\bibnamefont {van~der
  Sar}}, \bibinfo {author} {\bibfnamefont {F.}~\bibnamefont {Jelezko}},
  \bibinfo {author} {\bibfnamefont {V.~V.}\ \bibnamefont {Dobrovitski}}, \ and\
  \bibinfo {author} {\bibfnamefont {R.}~\bibnamefont {Hanson}},\ }\bibfield
  {title} {\enquote {\bibinfo {title} {Detection and control of individual
  nuclear spins using a weakly coupled electron spin},}\ }\href {\doibase
  10.1103/PhysRevLett.109.137602} {\bibfield  {journal} {\bibinfo  {journal}
  {Physical Review Letters}\ }\textbf {\bibinfo {volume} {109}},\ \bibinfo
  {pages} {137602} (\bibinfo {year} {2012})}\BibitemShut {NoStop}%
\bibitem [{\citenamefont {Wang}\ \emph {et~al.}(2016)\citenamefont {Wang},
  \citenamefont {Haase}, \citenamefont {Casanova},\ and\ \citenamefont
  {Plenio}}]{wang_positioning_2016}%
  \BibitemOpen
  \bibfield  {author} {\bibinfo {author} {\bibfnamefont {Zhen-Yu}\ \bibnamefont
  {Wang}}, \bibinfo {author} {\bibfnamefont {Jan~F.}\ \bibnamefont {Haase}},
  \bibinfo {author} {\bibfnamefont {Jorge}\ \bibnamefont {Casanova}}, \ and\
  \bibinfo {author} {\bibfnamefont {Martin~B.}\ \bibnamefont {Plenio}},\
  }\bibfield  {title} {\enquote {\bibinfo {title} {Positioning nuclear spins in
  interacting clusters for quantum technologies and bioimaging},}\ }\href
  {\doibase 10.1103/PhysRevB.93.174104} {\bibfield  {journal} {\bibinfo
  {journal} {Phys. Rev. B}\ }\textbf {\bibinfo {volume} {93}},\ \bibinfo
  {pages} {174104} (\bibinfo {year} {2016})}\BibitemShut {NoStop}%
\bibitem [{\citenamefont {Zopes}\ \emph {et~al.}(2018)\citenamefont {Zopes},
  \citenamefont {Herb}, \citenamefont {Cujia},\ and\ \citenamefont
  {Degen}}]{zopes_three-dimensional_2018}%
  \BibitemOpen
  \bibfield  {author} {\bibinfo {author} {\bibfnamefont {J.}~\bibnamefont
  {Zopes}}, \bibinfo {author} {\bibfnamefont {K.}~\bibnamefont {Herb}},
  \bibinfo {author} {\bibfnamefont {K.~S.}\ \bibnamefont {Cujia}}, \ and\
  \bibinfo {author} {\bibfnamefont {C.~L.}\ \bibnamefont {Degen}},\ }\bibfield
  {title} {\enquote {\bibinfo {title} {Three-dimensional nuclear spin
  positioning using coherent radio-frequency control},}\ }\href {\doibase
  10.1103/PhysRevLett.121.170801} {\bibfield  {journal} {\bibinfo  {journal}
  {Phys. Rev. Lett.}\ }\textbf {\bibinfo {volume} {121}},\ \bibinfo {pages}
  {170801} (\bibinfo {year} {2018})}\BibitemShut {NoStop}%
\bibitem [{\citenamefont {Abobeih}\ \emph {et~al.}(2019)\citenamefont
  {Abobeih}, \citenamefont {Randall}, \citenamefont {Bradley}, \citenamefont
  {Bartling}, \citenamefont {Bakker}, \citenamefont {Degen}, \citenamefont
  {Markham}, \citenamefont {Twitchen},\ and\ \citenamefont
  {Taminiau}}]{abobeih_atomic-scale_2019}%
  \BibitemOpen
  \bibfield  {author} {\bibinfo {author} {\bibfnamefont {M.~H.}\ \bibnamefont
  {Abobeih}}, \bibinfo {author} {\bibfnamefont {J.}~\bibnamefont {Randall}},
  \bibinfo {author} {\bibfnamefont {C.~E.}\ \bibnamefont {Bradley}}, \bibinfo
  {author} {\bibfnamefont {H.~P.}\ \bibnamefont {Bartling}}, \bibinfo {author}
  {\bibfnamefont {M.~A.}\ \bibnamefont {Bakker}}, \bibinfo {author}
  {\bibfnamefont {M.~J.}\ \bibnamefont {Degen}}, \bibinfo {author}
  {\bibfnamefont {M.}~\bibnamefont {Markham}}, \bibinfo {author} {\bibfnamefont
  {D.~J.}\ \bibnamefont {Twitchen}}, \ and\ \bibinfo {author} {\bibfnamefont
  {T.~H.}\ \bibnamefont {Taminiau}},\ }\bibfield  {title} {\enquote {\bibinfo
  {title} {Atomic-scale imaging of a 27-nuclear-spin cluster using a quantum
  sensor},}\ }\href {\doibase 10.1038/s41586-019-1834-7} {\bibfield  {journal}
  {\bibinfo  {journal} {Nature}\ }\textbf {\bibinfo {volume} {576}},\ \bibinfo
  {pages} {411--415} (\bibinfo {year} {2019})}\BibitemShut {NoStop}%
\bibitem [{\citenamefont {van~de Stolpe}\ \emph {et~al.}(2024)\citenamefont
  {van~de Stolpe}, \citenamefont {Kwiatkowski}, \citenamefont {Bradley},
  \citenamefont {Randall}, \citenamefont {Abobeih}, \citenamefont
  {Breitweiser}, \citenamefont {Bassett}, \citenamefont {Markham},
  \citenamefont {Twitchen},\ and\ \citenamefont
  {Taminiau}}]{van_de_stolpe_mapping_2024}%
  \BibitemOpen
  \bibfield  {author} {\bibinfo {author} {\bibfnamefont {G.~L.}\ \bibnamefont
  {van~de Stolpe}}, \bibinfo {author} {\bibfnamefont {D.~P.}\ \bibnamefont
  {Kwiatkowski}}, \bibinfo {author} {\bibfnamefont {C.~E.}\ \bibnamefont
  {Bradley}}, \bibinfo {author} {\bibfnamefont {J.}~\bibnamefont {Randall}},
  \bibinfo {author} {\bibfnamefont {M.~H.}\ \bibnamefont {Abobeih}}, \bibinfo
  {author} {\bibfnamefont {S.~A.}\ \bibnamefont {Breitweiser}}, \bibinfo
  {author} {\bibfnamefont {L.~C.}\ \bibnamefont {Bassett}}, \bibinfo {author}
  {\bibfnamefont {M.}~\bibnamefont {Markham}}, \bibinfo {author} {\bibfnamefont
  {D.~J.}\ \bibnamefont {Twitchen}}, \ and\ \bibinfo {author} {\bibfnamefont
  {T.~H.}\ \bibnamefont {Taminiau}},\ }\bibfield  {title} {{\selectlanguage
  {english}\enquote {\bibinfo {title} {Mapping a 50-spin-qubit network through
  correlated sensing},}\ }}\href {\doibase 10.1038/s41467-024-46075-4}
  {\bibfield  {journal} {\bibinfo  {journal} {Nature Communications}\ }\textbf
  {\bibinfo {volume} {15}},\ \bibinfo {pages} {2006} (\bibinfo {year}
  {2024})},\ \bibinfo {note} {publisher: Nature Publishing Group}\BibitemShut
  {NoStop}%
\bibitem [{\citenamefont {Szász}\ \emph {et~al.}(2013)\citenamefont {Szász},
  \citenamefont {Hornos}, \citenamefont {Marsman},\ and\ \citenamefont
  {Gali}}]{szasz_hyperfine_2013}%
  \BibitemOpen
  \bibfield  {author} {\bibinfo {author} {\bibfnamefont {Krisztián}\
  \bibnamefont {Szász}}, \bibinfo {author} {\bibfnamefont {Tamás}\
  \bibnamefont {Hornos}}, \bibinfo {author} {\bibfnamefont {Martijn}\
  \bibnamefont {Marsman}}, \ and\ \bibinfo {author} {\bibfnamefont {Adam}\
  \bibnamefont {Gali}},\ }\bibfield  {title} {\enquote {\bibinfo {title}
  {Hyperfine coupling of point defects in semiconductors by hybrid density
  functional calculations: The role of core spin polarization},}\ }\href
  {\doibase 10.1103/PhysRevB.88.075202} {\bibfield  {journal} {\bibinfo
  {journal} {Physical Review B}\ }\textbf {\bibinfo {volume} {88}},\ \bibinfo
  {pages} {075202} (\bibinfo {year} {2013})}\BibitemShut {NoStop}%
\bibitem [{\citenamefont {Son}\ \emph {et~al.}()\citenamefont {Son},
  \citenamefont {Carlsson}, \citenamefont {ul~Hassan}, \citenamefont {Janzén},
  \citenamefont {Umeda}, \citenamefont {Isoya}, \citenamefont {Gali},
  \citenamefont {Bockstedte}, \citenamefont {Morishita}, \citenamefont
  {Ohshima},\ and\ \citenamefont {Itoh}}]{son_divacancy_2006}%
  \BibitemOpen
  \bibfield  {author} {\bibinfo {author} {\bibfnamefont {N.~T.}\ \bibnamefont
  {Son}}, \bibinfo {author} {\bibfnamefont {P.}~\bibnamefont {Carlsson}},
  \bibinfo {author} {\bibfnamefont {J.}~\bibnamefont {ul~Hassan}}, \bibinfo
  {author} {\bibfnamefont {E.}~\bibnamefont {Janzén}}, \bibinfo {author}
  {\bibfnamefont {T.}~\bibnamefont {Umeda}}, \bibinfo {author} {\bibfnamefont
  {J.}~\bibnamefont {Isoya}}, \bibinfo {author} {\bibfnamefont
  {A.}~\bibnamefont {Gali}}, \bibinfo {author} {\bibfnamefont {M.}~\bibnamefont
  {Bockstedte}}, \bibinfo {author} {\bibfnamefont {N.}~\bibnamefont
  {Morishita}}, \bibinfo {author} {\bibfnamefont {T.}~\bibnamefont {Ohshima}},
  \ and\ \bibinfo {author} {\bibfnamefont {H.}~\bibnamefont {Itoh}},\
  }\bibfield  {title} {\enquote {\bibinfo {title} {Divacancy in 4h-{SiC}},}\
  }\href {\doibase 10.1103/PhysRevLett.96.055501} {\bibfield  {journal}
  {\bibinfo  {journal} {Phys. Rev. Lett.}\ }\textbf {\bibinfo {volume} {96}},\
  \bibinfo {pages} {055501}}\BibitemShut {NoStop}%
\bibitem [{\citenamefont {Isoya}\ \emph {et~al.}()\citenamefont {Isoya},
  \citenamefont {Umeda}, \citenamefont {Mizuochi}, \citenamefont {Son},
  \citenamefont {Janzén},\ and\ \citenamefont {Ohshima}}]{isoya_epr_2008}%
  \BibitemOpen
  \bibfield  {author} {\bibinfo {author} {\bibfnamefont {J.}~\bibnamefont
  {Isoya}}, \bibinfo {author} {\bibfnamefont {T.}~\bibnamefont {Umeda}},
  \bibinfo {author} {\bibfnamefont {N.}~\bibnamefont {Mizuochi}}, \bibinfo
  {author} {\bibfnamefont {N.~T.}\ \bibnamefont {Son}}, \bibinfo {author}
  {\bibfnamefont {E.}~\bibnamefont {Janzén}}, \ and\ \bibinfo {author}
  {\bibfnamefont {T.}~\bibnamefont {Ohshima}},\ }\bibfield  {title} {\enquote
  {\bibinfo {title} {{EPR} identification of intrinsic defects in {SiC}},}\
  }\href {\doibase 10.1002/pssb.200844209} {\bibfield  {journal} {\bibinfo
  {journal} {physica status solidi (b)}\ }\textbf {\bibinfo {volume} {245}},\
  \bibinfo {pages} {1298--1314}}\BibitemShut {NoStop}%
\bibitem [{\citenamefont {Ivády}\ \emph {et~al.}(2018)\citenamefont {Ivády},
  \citenamefont {Abrikosov},\ and\ \citenamefont {Gali}}]{ivady_first_2018}%
  \BibitemOpen
  \bibfield  {author} {\bibinfo {author} {\bibfnamefont {Viktor}\ \bibnamefont
  {Ivády}}, \bibinfo {author} {\bibfnamefont {Igor~A.}\ \bibnamefont
  {Abrikosov}}, \ and\ \bibinfo {author} {\bibfnamefont {Adam}\ \bibnamefont
  {Gali}},\ }\bibfield  {title} {\enquote {\bibinfo {title} {First principles
  calculation of spin-related quantities for point defect qubit research},}\
  }\href {\doibase 10.1038/s41524-018-0132-5} {\bibfield  {journal} {\bibinfo
  {journal} {npj Computational Materials}\ }\textbf {\bibinfo {volume} {4}},\
  \bibinfo {pages} {1--13} (\bibinfo {year} {2018})}\BibitemShut {NoStop}%
\bibitem [{\citenamefont {Nizovtsev}\ \emph {et~al.}(2014)\citenamefont
  {Nizovtsev}, \citenamefont {Kilin}, \citenamefont {Pushkarchuk},
  \citenamefont {Pushkarchuk},\ and\ \citenamefont
  {Jelezko}}]{nizovtsev_theoretical_2014}%
  \BibitemOpen
  \bibfield  {author} {\bibinfo {author} {\bibfnamefont {A.~P.}\ \bibnamefont
  {Nizovtsev}}, \bibinfo {author} {\bibfnamefont {S.~Ya}\ \bibnamefont
  {Kilin}}, \bibinfo {author} {\bibfnamefont {A.~L.}\ \bibnamefont
  {Pushkarchuk}}, \bibinfo {author} {\bibfnamefont {V.~A.}\ \bibnamefont
  {Pushkarchuk}}, \ and\ \bibinfo {author} {\bibfnamefont {F.}~\bibnamefont
  {Jelezko}},\ }\bibfield  {title} {\enquote {\bibinfo {title} {Theoretical
  study of hyperfine interactions and optically detected magnetic resonance
  spectra by simulation of the c291[{NV}]-h172diamond cluster hosting
  nitrogen-vacancy center},}\ }\href {\doibase 10.1088/1367-2630/16/8/083014}
  {\bibfield  {journal} {\bibinfo  {journal} {New Journal of Physics}\ }\textbf
  {\bibinfo {volume} {16}},\ \bibinfo {pages} {083014} (\bibinfo {year}
  {2014})}\BibitemShut {NoStop}%
\bibitem [{\citenamefont {Nizovtsev}\ \emph {et~al.}()\citenamefont
  {Nizovtsev}, \citenamefont {Kilin}, \citenamefont {Pushkarchuk},
  \citenamefont {Pushkarchuk}, \citenamefont {Kuten}, \citenamefont {Zhikol},
  \citenamefont {Schmitt}, \citenamefont {Unden},\ and\ \citenamefont
  {Jelezko}}]{nizovtsev_non-flipping13c_2018}%
  \BibitemOpen
  \bibfield  {author} {\bibinfo {author} {\bibfnamefont {A.~P.}\ \bibnamefont
  {Nizovtsev}}, \bibinfo {author} {\bibfnamefont {S.~Ya}\ \bibnamefont
  {Kilin}}, \bibinfo {author} {\bibfnamefont {A.~L.}\ \bibnamefont
  {Pushkarchuk}}, \bibinfo {author} {\bibfnamefont {V.~A.}\ \bibnamefont
  {Pushkarchuk}}, \bibinfo {author} {\bibfnamefont {S.~A.}\ \bibnamefont
  {Kuten}}, \bibinfo {author} {\bibfnamefont {O.~A.}\ \bibnamefont {Zhikol}},
  \bibinfo {author} {\bibfnamefont {S.}~\bibnamefont {Schmitt}}, \bibinfo
  {author} {\bibfnamefont {T.}~\bibnamefont {Unden}}, \ and\ \bibinfo {author}
  {\bibfnamefont {F.}~\bibnamefont {Jelezko}},\ }\bibfield  {title} {\enquote
  {\bibinfo {title} {Non-flipping 13c spins near an {NV} center in diamond:
  hyperfine and spatial characteristics by density functional theory simulation
  of the c510[{NV}]h252cluster},}\ }\href {\doibase 10.1088/1367-2630/aaa910}
  {\bibfield  {journal} {\bibinfo  {journal} {New Journal of Physics}\ }\textbf
  {\bibinfo {volume} {20}},\ \bibinfo {pages} {023022}}\BibitemShut {NoStop}%
\bibitem [{\citenamefont {Kresse}\ and\ \citenamefont {Hafner}(1994)}]{VASP}%
  \BibitemOpen
  \bibfield  {author} {\bibinfo {author} {\bibfnamefont {G.}~\bibnamefont
  {Kresse}}\ and\ \bibinfo {author} {\bibfnamefont {J.}~\bibnamefont
  {Hafner}},\ }\bibfield  {title} {\enquote {\bibinfo {title} {Ab initio
  molecular-dynamics simulation of the liquid-metal–amorphous-semiconductor
  transition in germanium},}\ }\href {\doibase 10.1103/PhysRevB.49.14251}
  {\bibfield  {journal} {\bibinfo  {journal} {Phys. Rev. B}\ }\textbf {\bibinfo
  {volume} {49}},\ \bibinfo {pages} {14251--14269} (\bibinfo {year}
  {1994})}\BibitemShut {NoStop}%
\bibitem [{\citenamefont {Kresse}\ and\ \citenamefont
  {Furthm\"uller}(1996)}]{VASP2}%
  \BibitemOpen
  \bibfield  {author} {\bibinfo {author} {\bibfnamefont {G.}~\bibnamefont
  {Kresse}}\ and\ \bibinfo {author} {\bibfnamefont {J.}~\bibnamefont
  {Furthm\"uller}},\ }\bibfield  {title} {\enquote {\bibinfo {title} {Efficient
  iterative schemes for \textit{ab initio} total-energy calculations using a
  plane-wave basis set},}\ }\href {\doibase 10.1103/PhysRevB.54.11169}
  {\bibfield  {journal} {\bibinfo  {journal} {Phys. Rev. B}\ }\textbf {\bibinfo
  {volume} {54}},\ \bibinfo {pages} {11169--11186} (\bibinfo {year}
  {1996})}\BibitemShut {NoStop}%
\bibitem [{\citenamefont {Heyd}\ \emph {et~al.}(2003)\citenamefont {Heyd},
  \citenamefont {Scuseria},\ and\ \citenamefont
  {Ernzerhof}}]{heyd_hybrid_2003}%
  \BibitemOpen
  \bibfield  {author} {\bibinfo {author} {\bibfnamefont {Jochen}\ \bibnamefont
  {Heyd}}, \bibinfo {author} {\bibfnamefont {Gustavo~E.}\ \bibnamefont
  {Scuseria}}, \ and\ \bibinfo {author} {\bibfnamefont {Matthias}\ \bibnamefont
  {Ernzerhof}},\ }\bibfield  {title} {\enquote {\bibinfo {title} {Hybrid
  functionals based on a screened coulomb potential},}\ }\href {\doibase
  10.1063/1.1564060} {\ \textbf {\bibinfo {volume} {118}},\ \bibinfo {pages}
  {8207--8215} (\bibinfo {year} {2003})}\BibitemShut {NoStop}%
\bibitem [{\citenamefont {Heyd}\ \emph {et~al.}(2006)\citenamefont {Heyd},
  \citenamefont {Scuseria},\ and\ \citenamefont
  {Ernzerhof}}]{heyd_erratum:_2006}%
  \BibitemOpen
  \bibfield  {author} {\bibinfo {author} {\bibfnamefont {Jochen}\ \bibnamefont
  {Heyd}}, \bibinfo {author} {\bibfnamefont {Gustavo~E.}\ \bibnamefont
  {Scuseria}}, \ and\ \bibinfo {author} {\bibfnamefont {Matthias}\ \bibnamefont
  {Ernzerhof}},\ }\bibfield  {title} {\enquote {\bibinfo {title} {Erratum:
  “hybrid functionals based on a screened coulomb potential” [ j. chem.
  phys. 118, 8207 (2003) ]},}\ }\href {\doibase 10.1063/1.2204597} {\ \textbf
  {\bibinfo {volume} {124}},\ \bibinfo {pages} {219906} (\bibinfo {year}
  {2006})}\BibitemShut {NoStop}%
\bibitem [{Note1()}]{Note1}%
  \BibitemOpen
  \bibinfo {note} {Theoretical hyperfine tensors for all lattice sites within
  30~\r A\ distance from the NV center are available at \protect \href
  {https://ivadygroup.elte.hu/hyperfine}{https://ivadygroup.elte.hu/hyperfine}\label
  {footnote1}}\BibitemShut {NoStop}%
\bibitem [{\citenamefont {Blöchl}(2000)}]{blochl_first-principles_2000}%
  \BibitemOpen
  \bibfield  {author} {\bibinfo {author} {\bibfnamefont {Peter~E.}\
  \bibnamefont {Blöchl}},\ }\bibfield  {title} {\enquote {\bibinfo {title}
  {First-principles calculations of defects in oxygen-deficient silica exposed
  to hydrogen},}\ }\href {\doibase 10.1103/PhysRevB.62.6158} {\bibfield
  {journal} {\bibinfo  {journal} {Physical Review B}\ }\textbf {\bibinfo
  {volume} {62}},\ \bibinfo {pages} {6158--6179} (\bibinfo {year}
  {2000})}\BibitemShut {NoStop}%
\bibitem [{\citenamefont {Bl\"ochl}(1994)}]{PAW}%
  \BibitemOpen
  \bibfield  {author} {\bibinfo {author} {\bibfnamefont {P.~E.}\ \bibnamefont
  {Bl\"ochl}},\ }\bibfield  {title} {\enquote {\bibinfo {title} {Projector
  augmented-wave method},}\ }\href {\doibase 10.1103/PhysRevB.50.17953}
  {\bibfield  {journal} {\bibinfo  {journal} {Phys. Rev. B}\ }\textbf {\bibinfo
  {volume} {50}},\ \bibinfo {pages} {17953--17979} (\bibinfo {year}
  {1994})}\BibitemShut {NoStop}%
\bibitem [{\citenamefont {Yazyev}\ \emph {et~al.}(2005)\citenamefont {Yazyev},
  \citenamefont {Tavernelli}, \citenamefont {Helm},\ and\ \citenamefont
  {Röthlisberger}}]{yazyev_core_2005}%
  \BibitemOpen
  \bibfield  {author} {\bibinfo {author} {\bibfnamefont {Oleg~V.}\ \bibnamefont
  {Yazyev}}, \bibinfo {author} {\bibfnamefont {Ivano}\ \bibnamefont
  {Tavernelli}}, \bibinfo {author} {\bibfnamefont {Lothar}\ \bibnamefont
  {Helm}}, \ and\ \bibinfo {author} {\bibfnamefont {Ursula}\ \bibnamefont
  {Röthlisberger}},\ }\bibfield  {title} {\enquote {\bibinfo {title} {Core
  spin-polarization correction in pseudopotential-based electronic structure
  calculations},}\ }\href {\doibase 10.1103/PhysRevB.71.115110} {\bibfield
  {journal} {\bibinfo  {journal} {Physical Review B}\ }\textbf {\bibinfo
  {volume} {71}},\ \bibinfo {pages} {115110} (\bibinfo {year}
  {2005})}\BibitemShut {NoStop}%
\bibitem [{\citenamefont {Perdew}\ \emph {et~al.}(1996)\citenamefont {Perdew},
  \citenamefont {Burke},\ and\ \citenamefont {Ernzerhof}}]{PBE}%
  \BibitemOpen
  \bibfield  {author} {\bibinfo {author} {\bibfnamefont {John~P.}\ \bibnamefont
  {Perdew}}, \bibinfo {author} {\bibfnamefont {Kieron}\ \bibnamefont {Burke}},
  \ and\ \bibinfo {author} {\bibfnamefont {Matthias}\ \bibnamefont
  {Ernzerhof}},\ }\bibfield  {title} {\enquote {\bibinfo {title} {Generalized
  gradient approximation made simple},}\ }\href {\doibase
  10.1103/PhysRevLett.77.3865} {\bibfield  {journal} {\bibinfo  {journal}
  {Phys. Rev. Lett.}\ }\textbf {\bibinfo {volume} {77}},\ \bibinfo {pages}
  {3865--3868} (\bibinfo {year} {1996})}\BibitemShut {NoStop}%
\end{thebibliography}
%

\section*{Acknowledgments} 

Insightful comments from Alexander P.\ Nizovtsev are highly appreciated. This work was supported by the National Research, 
Development and Innovation Office of Hungary (NKFIH) within the Quantum Information National Laboratory of Hungary (Grant No. 2022-2.1.1-NL-2022-00004) and within project FK 145395. 
V.I. acknowledges support from the Knut and Alice Wallenberg Foundation through the WBSQD2 project (Grant No.\ 2018.0071). The computations were enabled by resources provided by the National Academic Infrastructure for Supercomputing in Sweden (NAISS) at the Swedish National Infrastructure for Computing (SNIC) at Tetralith, partially funded by the Swedish Research Council through grant agreement no. 2022-06725.

\section*{Author contributions} 

I.T. and V.I. carried out the first principles calculations, and V.I. developed and implemented the real-space integration code. V.I. and I.T. wrote the manuscript. The work was supervised by V.I. 

\section*{Competing interests} 

The authors declare no competing interests.

\end{document}